\documentclass[a4paper,11pt,fleqn]{article}
\usepackage{mathrsfs}

\begin{document}

%%%%%%%%%%%%%%%%%%%%% JOURNALS {no}{year}{page} %%%%%%%%%%%%%%%%%%%%%%
\def\ijmp#1#2#3{{\it Int.~J.~Mod.~Phys.\/}~{\bf A#1} (19#2) #3}
\def\ib#1#2#3{{\it ibid.\/}~{\bf#1} (19#2) #3}
\def\mpla#1#2#3{{\it Mod.~Phys.~Lett.\/}~{\bf A#1} (19#2) #3}
\def\np#1#2#3{{\it Nucl.~Phys.\/}~{\bf B#1} (19#2) #3}
\def\pl#1#2#3{{\it Phys.~Lett.\/}~{\bf B#1} (19#2) #3}
\def\pp{{\it preprint\/} }
\def\prd#1#2#3{{\it Phys.~Rev.\/}~{\bf D#1} (19#2) #3}
\def\prl#1#2#3{{\it Phys.~Rev.~Lett.\/}~{\bf #1} (19#2) #3}
\def\prp#1#2#3{{\it Phys.~Rep.\/}~{\bf #1} (19#2) #3}
\def\ptp#1#2#3{{\it Prog.~Theor.~Phys.\/}~{\bf #1} (19#2) #3}
\def\rmp#1#2#3{{\it Rev.~Mod.~Phys.\/}~{\bf #1} (19#2) #3}
\def\zpc#1#2#3{{\it Z.~Phys.\/}~{\bf C#1} (19#2) #3}
\def\arn#1#2#3{{\it Ann.~Rev.~Nucl.~Part.~Sci.\/}~{\bf #1} (19#2) #3}
\def\Np#1#2#3{{\it Nucl.~Phys.\/}~{\bf#1} (19#2) #3}
\def\pr#1#2#3{{\it Phys.~Rep.\/}~{\bf #1} (19#2) #3}
\def\jhep#1#2#3{{\it JHEP\/}~{\bf #1} (19#2) #3}
%%%%%%%%%%%%%%%%%%%%%%%%%%%%%%%%%%%%%%%%%%%%%%%%%%%%%%%%%%%%%%%%%%%%%% 

 \def\Ca{C{\!_{\!A}}}
 \def\Cf{C{\!_F}}
 \def\Nf{N{\!_f}}
 \def\erf{\mbox{erf}}
 \def\epem{\ensuremath{e^{+}e^{-}\,}}
 \def\epemgt{\epem~$ \rightarrow\!$~}
 \def\qt{\tilde{q}}
 \def\R4{$R_4(y_{cut})$}
 \def\B0{\ensuremath{\beta_0}}
 \def\as{\ensuremath{\alpha_s}}
 \def\a{\ensuremath{\frac{\as}{\pi}}}
 \def\rd{\mathrm{d}}
 \def\ycut{\ensuremath{y_{cut}}}
 \newcommand{\lf}[2] {\mbox{$\frac{#1}{#2}$}}
 \newcommand{\Gq}[2] {\ensuremath{\Gamma_{q}(#1,#2)}}
 \newcommand{\Gg}[2] {\ensuremath{\Gamma_{g}(#1,#2)}} 
 \newcommand{\Gf}[1] {\ensuremath{\Gamma_{f}(#1)}}
 \newcommand{\Dq}[1] {\ensuremath{\Delta_{q}(#1)}}
 \newcommand{\Dg}[1] {\ensuremath{\Delta_{g}(#1)}}
 \newcommand{\Df}[1] {\ensuremath{\Delta_{f}(#1)}}
 \newcommand{\sss}{\scriptscriptstyle}
 \newcommand{\fanC}[1] {\ensuremath{\mathscr{C}_{\sss{#1}}}}
 
%%% Local Variables: 
%%% mode: plain-tex
%%% TeX-master: t
%%% End: 
 
\renewcommand{\thefootnote}{\fnsymbol{footnote}}
\begin{flushright}  
%hep-ph/9902305\\
DTP/99/12\\
February 1999\\  
\end{flushright}
\nopagebreak
\vspace{0.75cm}
\begin{center}
\LARGE
{\bf The Four-Jet Rate in e$^{+}$e$^{-}$ Annihilation}
\vspace{0.6cm}
\Large

Stephen~J.~Burby\footnote{Email: \texttt{s.j.burby@dur.ac.uk}}

\vspace{0.4cm}
\large
\begin{em}
Centre for Particle Theory, University of Durham\\
South Road, Durham, DH1 3LE, England
\end{em}

\vspace{1.7cm}

\end{center}
\normalsize
\vspace{0.45cm}
%%%%%%%%%%%%ABSTRACT%%%%%%%%%%%%%%%%%%
\begin{abstract}  
  We present an analytic expression for the four-jet rate (\(R_{4}\)) in
  \epem annihilation calculated using the coherent branching formalism in the
  \(k_\bot \)\,(Durham) scheme. Our result resums all the leading and
  next-to-leading kinematic logarithms to all orders in the QCD strong
  coupling constant \as.
\end{abstract}

\newpage 

\section*{Introduction}

The study of \epem annihilation has always been one of the most useful means
of exploring QCD and determining a value for $\as(M_z)$, the strong coupling
constant. Multi-jet rates\footnote{The $n$-jet rate is defined as $R_{n}(Q) =
  \sigma_{n\textrm{-}jet}/\sigma_{hadrons}$} in particular enable us to
examine the perturbative nature of QCD with long distance effects kept
comparatively low.  It is also important to make sure that the shape variable
being investigated is \textit{both} collinear \textit{and} infra-red safe.
Jet rates can satisfy all of these criteria as long as care is taken in the
choice of the jet clustering algorithm. In this paper we calculate the
leading and next-to-leading logarithmic contribution to the four-jet rate for
\epem annihilation using the Durham algorithm. (For an explanation of various
algorithms and the reasoning behind choosing the Durham one see
refs.~\cite{durham,brown,jetalg}). We then obtain an expression for the jet
rate in terms of a dimensionless jet resolution parameter, \ycut, which can
be considered as a measure of how well we are able to resolve two
approximately collinear partons. According to the Durham algorithm we define
$\ycut=Q_0^2/Q^2$, where $Q \sim \sqrt{s}$ is the scale of the jet-production
process and hence the cut-off energy scale $Q_0$ can be considered to be the
energy threshold below which the process starts to become non-perturbative.

In the region of small \ycut ($\ll1$) the emitted gluons are predominantly
soft and collinear resulting in the logarithmic enhancement of higher orders
\cite{muel,catani}. It is therefore necessary to resum them to all orders in
\as\mbox{} to obtain a reliable prediction for the four-jet rate.
  
\subsection*{Leading Logarithms and Exponentiation}
First it is important to stress what we are actually calculating in the
resummation procedure. Using the coherent branching formalism
\cite{catani,pQCD,fadin}, we are able to resum \textit{exactly} all
contributions to the shape variable at leading logarithms, LL ($\as^n
L^{2n}$) and next-to-leading logarithms, NLL ($\as^n L^{2n-1}$) in the
perturbative expansion where $L=-\ln(\ycut)$. This means that all terms
sub-leading are not completely reproduced and therefore they are dropped in
our calculation.

The idea behind exponentiation is to increase the domain of applicability of
the shape variable such that it extends into the region of $\as L \leq 1$.
The result of this procedure is to obtain a closed function of the form
$\mathcal{F}\left(L g_1(\as L) + g_2(\as L)\right)$, where $g_1(\as L)$
resums all leading-logarithmic contributions and $g_2(\as L)$ resums the
next-to-leading ones such that when expanded the whole perturbation series is
reproduced down to terms of the form $\as^n L^m $, where $n\le m\le 2n$.  For
the jet fractions being studied, a simple exponentiation does not arise.  It
therefore only makes sense to calculate the LL and NLL contributions of the 
perturbation series.

\section*{Calculation}

\indent The primary aim of this paper is to find an analytic expression
for the $4$-jet rate, \(R_{4}(y_{cut}) \). The most simple way to solve this
is to work in terms of a generating function defined by  
\begin{equation}
\phi^{p}(Q,Q_{0};u)= \sum_{n=0}^{\infty}u^n R^p_n(Q,Q_0)
\end{equation}
where $R^p_n(Q,Q_0)$ is the probability of finding $n$-partons of a
particular type in the final state of a process, $p$, and $u$ is a jet label
to distinguish  each of the probabilities. In this case we are dealing with
\epemgt hadrons, therefore $\phi^{(\epem)} = [\phi_q]^2$, where
$\phi_q$ is the generating function for a single quark to branch,
\begin{equation}
\phi_q(Q,Q_0;u)=u+\int\limits^{\sss Q}_{\sss Q_{0}} \,
  \frac{\rd \qt}{\qt} \, \int\limits^{\sss
  1}_{\sss {Q_0/\tilde{q}}}\rd z \,
  \as (z\tilde{q})\,\frac{\Cf}{\pi} \left( \frac{2}{z}-\frac{3}{2}
  \right)\,[\phi_g(z\tilde{q},Q_0;u)-1]\, .
\end{equation}
 To obtain the $n$-jet rate, $R_n$, is simply a matter of
differentiating the generating function $n$ times at $u=0$.  The $n$-jet rate
is then
\begin{equation}
R_{n}(y_{cut}) = \frac{1}{n!}\left(\frac{\partial}{\partial
    u}\right)^{\!n}\! \left[ \phi_q(Q,Q_{0};u) \right]^2 \biggr|_{u=0}.
\end{equation} 
We find from the application of the coherent branching formalism, the
generating function obeys the following implicit coupled equations \cite{durham}:
\begin{equation}
\phi_{q}(Q,Q_{0};u) = u^{2}\exp\left(2\int_{Q_{0}}^{Q} \rd q\,
\Gamma_{q}(Q,q)[\phi_{g}(q,Q_{0};u) -1]\right)
\end{equation}
and 
\begin{eqnarray}
\phi_{g}(Q,Q_{0};u) =
 u\,\exp\biggr(\int_{Q_{0}}^{Q}\rd q
\{\Gamma_{g}(Q,q)[\phi_{g}(q,Q_{0};u)-1] -\Gamma_{f}(q)\} \biggr) \quad \quad
 \qquad \qquad \qquad \qquad\nonumber \\
\times \biggr(1+u\int_{Q_{0}}^{Q}\rd q\,\Gamma_{f}(q) 
\exp\biggr( \int_{Q_{0}}^{q}\rd q'\{[2\Gamma_{q}(q,q')-\Gamma_{g}(q,q')]
[\phi_{g}(q',Q_{0};u)-1]+\Gamma_{f}(q')\}\biggr)\biggr) .
\end{eqnarray}
Where the emission probabilities are defined as
\begin{eqnarray}
\Gq{Q}{q} &=& \frac{2\Cf}{\pi} \frac{\as(q)}{q} 
\left(\ln\frac{Q}{q}-\frac{3}{4} \right) , \\
\Gg{Q}{q} &=& \frac{2\Ca}{\pi} \frac{\as(q)}{q} 
\left(\ln\frac{Q}{q}-\frac{11}{12} \right) , \\
\Gf{q}\quad \,  &=& \frac{N_{f}}{3\pi} \frac{\as(q)}{q}.
\end{eqnarray}

The 2-jet limit is important as the jet rate becomes semi-inclusive
and exponentiation holds exactly. This gives 
\begin{eqnarray}
R_2(y_{cut}) = \exp\left(\frac{\Cf a L}{2} (3-L)-\B0\frac{\Cf a^2 L^3}{12}\right) ,
\end{eqnarray}
where $L=\ln(1/y_{cut})$, $a=\as(Q)/\pi$ and we have used $\B0=(11\Ca-2\Nf)/3$. The 3-jet case was evaluated
in \cite{charles} and we proceed in a similar way.

Firstly we find (3) gives, in the n=4 case \cite{durham},
\begin{eqnarray}
 R_4(y_{cut}) &=&  2 R_2(y_{cut}) \left( \int^{Q}_{Q_0} \rd q \, \Gq{Q}{q}\Dg{q}
 \int^{q}_{Q_0} \rd q' \, \Gg{q}{q'} \Dg{q'} \right) \nonumber \\
&&+ 2R_2(y_{cut}) \left(\int^{Q}_{Q_0} \rd q \, \Gq{Q}{q}\Dg{q}
 \int^{q}_{Q_0} \rd q' \, \Gf{q'} \Df{q'} \right)  \nonumber \\
&&+ R_3(y_{cut}) \left( \int^Q_{Q_0}\rd q \Gq{Q}{q} \Dg{q} \right) ,
\end{eqnarray}
where we have introduced the Sudakov form factors
\begin{eqnarray}
\Dq{Q} &=& \exp\left( -\int^{Q}_{Q_0}\rd q \,\Gq{Q}{q} \right), \\
\Dg{Q} &=& \exp\left( -\int^{Q}_{Q_0}\rd q \,[\,\Gg{Q}{q} +\Gf{q}] \right), \\
\Df{Q} &=& \exp\left( -\int^{Q}_{Q_0}\rd q 
\,[\,2\Gq{Q}{q} - \Gg{Q}{q} -\Gf{q}] \right). 
\end{eqnarray} 
We need only work with the one-loop definition of the strong coupling
constant,
\begin{equation}
\as(Q) = \frac{\as(\mu)}{1+\frac{\B0
      \as(\mu)}{2\pi}\ln\big({\frac{Q}{\mu}}\big)},
\end{equation} 
as higher order corrections will be sub-leading. Even at this order we are
still faced with an extremely complicated set of nested integrals.
Therefore, as in \cite{charles} we proceed by expressing \R4 as
\begin{equation}
R_4(y_{cut}) = R_4\biggr|_{\B0=0} + \B0 \frac{\partial
  R_4}{\partial \B0}\biggr|_{\B0=0}.
\end{equation}
This is permissable for any jet multiplicity evaluated at next-to-leading
logarithmic order because in general we will have 
\begin{eqnarray}
R_n &=& \fanC{12} a L^2 + \fanC{11} a L + \cdots \nonumber \\
      && + \fanC{24} a^2 L^4 + \fanC{23} a^2 L^3 +\cdots \nonumber \\
      && \vdots \nonumber \\
      &&+ \fanC{n\, 2n}a^n L^{2n} + \fanC{n\, 2n-1}a^n L^{2n-1} +\cdots,
\end{eqnarray}
where the coefficients \fanC{p\, q} are either \B0 independent (\fanC{p\,
  2p}) or contain a single \B0 (\fanC{p\, 2p-1}). All other \B0 dependence is
  contained in the strong coupling constant.
We note that
\begin{equation}
\frac{\partial [\as{\scriptstyle(Q)}]^m}{\partial \B0}\Biggr|_{\B0=0} =
-m \frac{[\as{\scriptstyle(Q)}]^{m+1}}{2\pi}\ln \!
{\textstyle\left(\frac{Q}{Q_0}\right)} \sim a^{m+1}L  .
\end{equation}
It is now apparent that beyond the first derivative there will only be terms
of the form $a^{n}L^{2n-2}$ which in the NLL approximation can be dropped. The
assumption of (15) then seems to be a valid one. In fact this expansion
greatly simplifies the calculation by enabling us to work with terms
evaluated with \B0 equal to zero. In doing so the coupling \as \, can really
be treated as a constant and hence no longer depends on the integration
variable. Proceeding in this way, we then calculate the four jet rate to be
\begin{eqnarray*}
  R_4&=&\lf{{\Cf^2}}{{\Ca^2}}\Big({e^{-2 (A+F)}} \big({{({e^A}-1)}^2} (2+3
   \Cf a L)  \\ & &
\hspace{1cm} -{\sqrt{\Ca a}}\, {e^A} ({e^A}-1) (-3+12 F+2 L) {\sqrt{\pi }}\, \mbox{erf}\big({\sqrt{A}}\big)\big)\Big)  \\ & & +
\lf\Cf\Ca\Big(\lf{1}{24} {e^{-2 F}} \big(-24 {e^{-2 A}} ({e^A}-1) (2+3 a \Cf  L)  \\ & & 
\hspace{1cm} -4 {\sqrt{\Ca a}}\, {e^{-A}} (2+3 {e^A} (-3+12 F+2 L)) {\sqrt{\pi }}\, \mbox{erf}\big({\sqrt{A}}\big)  \\ & & 
\hspace{1cm} -(12+a L (11 \Ca -6  \Cf  (-9+12 F+2 L))) \pi\, {{\mbox{erf}\big({\sqrt{A}}\big)}^2}  \\ & & 
\hspace{1cm} +2 {\sqrt{\Ca a}}
  (-7+72 F+12 L) {\sqrt{2 \pi }}\,
  \mbox{erf}\big({\sqrt{2A}}\big)\big)\Big)- \lf{11}{3}\,
  \mbox{{\Large$\varphi$}}  \\ & & 
+\lf{\beta_0 }\Ca \Bigg(\lf{{\Cf^2}}{{\Ca^2}}\lf{1}{12} \Big({e^{-2 (A+F)}} \big(-2 {\sqrt{\Ca a}} {e^A} (3+2 A (-3+4 F)) {\sqrt{\pi }}\, \mbox{erf}\big({\sqrt{A}}\big)\big)  \\ & & 
\hspace{1.5cm}\big(1+{e^A} \big(-1+{\sqrt{A}} {\sqrt{\pi }}\, \mbox{erf}\big ({\sqrt{A}}\big)\big)\big)\Big)  \\ & & 
\hspace{1cm}+\lf\Cf\Ca\lf{1}{12}\Big( {e^{-2 (A+F)}}2 \Cf a L (1+2 {e^A}+4 ({e^A}-1)F) \\ & & 
\hspace{1.5cm} \big(1+{e^A} \big(-1+{\sqrt{A}} {\sqrt{\pi }}\,
   \mbox{erf}\big({\sqrt{A}}\big)\big)\big) \Big)  \\ & & 
\hspace{1cm}+{\sqrt{\lf\Cf\Ca}}\lf{1}{24} \Big({\sqrt{\Cf a}}\, {e^{-2
   (A+F)}} \big(2 {e^A} (5+{e^A} \\ & & 
\hspace{1.5cm}+ A (-4-2{e^A}(3-8F))) {\sqrt{\pi }}\, \mbox{erf}\big({\sqrt{A}}\big)  \\ & & 
\hspace{1.5cm}- {e^{2 A}}(9-4A(3-8F)) {\sqrt{2 \pi }}\,
   \mbox{erf}\big({\sqrt{2A}}\big)\big) \Big)  \\ & & 
\hspace{1cm}+\lf{1}{12} \Cf aL\, {e^{-2 (A+F)}} (1-{e^A}(1-3{e^A})-8F(1-{e^A})\\&&
\hspace{1cm}+2F{e^{2A}} \pi {\mbox{erf\,}^2}\big({\sqrt{A}}\big)) + \mbox{{\Large$\varphi$}} \Bigg), \\
\end{eqnarray*}
erf$(x)$ is the error function defined to be $\frac{2}{\sqrt{\pi}}\int^x_0
e^{-y^2} \rd y$ and erf\mbox{}i$(x) =$ erf$(i x)/i$.  We have also defined
$A=\Ca aL^2/4,\ F=\Cf aL^2/4$ and {\large$\mathcal{G}$}$(x,z)= x\int
_{0}^{z}{e^{x {y^2}}} \mbox{erf}(y)\rd y$. Attempts were made to solve
{\large$\mathcal{G}$} exactly but no closed form was found. It appears that
the integral is just a generalisation of the error function and hence cannot
be solved, except in certain cases. A reference containing various properties
of this function is given \cite{rosser}.

\newpage
\subsection*{Properties of the Four-Jet Rate}
With the complete result calculated we are able to reproduce the exact LL and
NLL coefficients of \as\mbox{} at any order. The first three orders in
\as/$\pi$ are
given.  

i.e.
$R_4(y_{cut})=a^2(B_4L^4+B_3L^3+\mathcal{O}(L^2))+
a^3(C_6L^6+C_5L^5+\mathcal{O}(L^4))+
a^4(D_8L^8+D_7L^7 +\mathcal{O}(L^6))+
\cdots$.
\begin{eqnarray*}
B_4&=& \lf{1}{8} \Cf^2+\lf{1}{48}\Cf\Ca . \\
%CORRECTION IN LINE BELOW. WAS \lf{-1}{4}, SHOULD BE \lf{-3}{4}
B_3&=& \lf{-3}{4}\Cf^2-\lf{5}{18}\Cf\Ca+\lf{1}{36}\Cf N_f. \\
C_6&=& \lf{-1}{16}\Cf^3 -\lf{1}{48}\Cf^2\Ca -\lf{7}{2880}\Cf\Ca\!^2. \\
C_5&=& \lf{9}{16}\Cf^3+\lf{71}{144}\Cf^2\Ca +\lf{217}{2880}\Cf\Ca\!^2
-\lf{41}{720}\Cf^2N_f-\lf{1}{120} \Ca\Cf N_f . \\
D_8&=& \lf{1}{64}\Cf^4 +\lf{1}{128}\Cf^3\Ca +\lf{1}{512}\Cf^2\Ca\!^2 
+\lf{1}{5120}\Cf\Ca\!^3. \\
%CORRECTION IN LINE BELOW. This was previously wrong in two terms
D_7&=& \lf{-3}{16}\Cf^4 -\lf{17}{64}\Cf^3\Ca -\lf{1439}{17280}\Cf^2\Ca\!^2 
-\lf{2371}{241920}\Cf\Ca\!^3 +\lf{323}{10080}\Cf^3 N_f
+\lf{31}{3024}\Cf^2\Ca N_f \\
&& +\lf{1}{840}\Cf\Ca\!^2 N_f. \\
\end{eqnarray*}
This is in agreement with \cite{durham} which gives the $B_{4,3}$
coefficients.  The $C_{6,5}$ coefficients were in addition calculated by
expanding out the integral equation (10) as a function of \as.  Another test
was to calculate $R_4$ in the large $N_c$ limit ($N_c$ is the number of
colours) to the order of leading logarithms. This greatly simplifies the
equations as $\Ca \rightarrow N_c$, $\Cf \rightarrow N_c/2$ and $N_f$ can be disregarded. Eq.~(4) now collapses down to
\begin{equation}
\phi(Q,Q_{0};u) = u^{2}\exp\left(2\int_{Q_{0}}^{Q} \rd q\, 
\Gamma_{q}(Q,q)\left[\frac{1}{u} \phi(q,Q_{0};u) -1\right]\right).
\end{equation}
Also noting that at leading logarithmic order $R_4$ will be independent of
$\beta_0$ we can safely set it to zero. We then get
\begin{eqnarray}
R_4^{N_c}= \frac{1}{4}e^{-3A} \left(6\, -\, 8e^A+\, 4\sqrt{A}\,
e^A(1-2e^A)\sqrt{\pi}\, \erf\sqrt{A}\, -\, (1-2A)\, e^{2A}\, \erf\,
^2\sqrt{A}\,  \right. \nonumber \\
\left. +2e^{2A}(1+2\sqrt{A}\, \sqrt{2\pi}\, \erf\sqrt{2A}) \right).
\end{eqnarray}
This is in agreement with the full NLL result in the appropriate limit.

We also note that in the psuedo-abelian limit of simply $\Ca$ and $N_f$ going
to zero, exact exponentiation holds. We also find that this gives a
reasonably good approximation to the full non-abelian case within about
15-20\%. 

\newpage
\section*{Conclusion}
To conclude, we have found an analytic expression for \R4 which exactly
resums the leading and next-to-leading logarithmic contributions to all
orders in \as. Previous work on including the leading and next-to-leading
logarithms to all orders \cite{signer} was performed by solving (10)
numerically. We propose that it is beneficial to have an explicit result
since it is then possible to extract the purely LL and NLL contributions and
drop the incomplete sub-leading terms.  Most importantly we will be able to
utilise the result to address the so called `renormalisation scale ambiguity'
in combining the resummed result with the fixed order one.  We will be
implementing a novel technique and hope to report on this in the future.

\subsection*{Acknowledgements}

I would firstly like to thank C.J.~Maxwell for suggesting the problem,
reading the manuscript and providing helpful comments. I am also grateful to
the U.K. Particle Physics and Astronomy Research Council, \mbox{(PPARC)}, for
a research~studentship.
Finally I would like to thank L.J.~Dixon for pointing out an error in the
original calculation and for useful subsequent conversations.

\end{document}